\documentclass[twocolumn,showpacs,showkeys]{revtex4}
\usepackage{graphicx}
\usepackage{bm}
\usepackage{color}
\usepackage{amsmath}
\usepackage{natbib}

\begin{document}

\title{Exchange Coulomb interaction in nanotubes: Dispersion of Langmuir waves}

\author{P. A. Andreev}%
\email{andreevpa@physics.msu.ru}

\author{A. Yu. Ivanov}%
\email{alexmax1989@mail.ru}

\affiliation{
Faculty of Physics, Lomonosov Moscow State University, Moscow, Russian
Federation.}

\date{\today}

\begin{abstract}
Microscopic derivation of the Coulomb exchange interaction for electrons located on the nanotubes is presented. Our derivation is based on the many-particle quantum hydrodynamic method. We demonstrate the role of the curvature of the nanocylinders on the force of the exchange interaction. We calculate corresponding dispersion dependencies for electron oscillations on the nanotubes.
\end{abstract}

\pacs{52.30.Ex, 73.63.Fg, 64.30.-t, 71.70.Gm}% PACS, the Physics and Astronomy
                             % Classification Scheme.
\keywords{quantum plasmas, exchange interaction, quantum hydrodynamics, 2DEG, nanotubes}
%Use showkeys class option if keyword

\maketitle

%52.30.Ex	Two-fluid and multi-fluid plasmas
%52.35.Dm	Sound waves

%52.27.Ep	Electron-positron plasmas

%64.30.-t	Equations of state of specific substances

%71.70.Gm Exchange interactions

%73.63.-b	Electronic transport in nanoscale materials and structures
%73.63.Fg	Nanotubes
%73.21.-b	Electron states and collective excitations in multilayers, quantum wells, mesoscopic, and nanoscale systems
%73.22.Lp	Collective excitations

\section{Introduction}

Studying of the Coulomb exchange interaction in electron gas has a long history \cite{Nozieres PR 58}-\cite{Brey PRB 90}. These fundamental parers are dedicated to the three dimensional electron gas. In 1983 the Coulomb exchange interaction was considered in the two dimensional electron gas by Datta \cite{Datta JAP 83}. It was performed on the background of research of the two dimensional electron gas in the self-consistent field approximation \cite{Krasheninnikov JETP 80}-\cite{Brown PRL 72}.

Generalization of the Coulomb exchange interaction force field in the three-dimensional electron gas (quantum plasmas) includes the dependence on the spin polarization \cite{MaksimovTMP 2001 b}, \cite{Andreev AtPhys 08}, \cite{Andreev 1403 exchange}. Same problem for the two dimensional quantum plasmas was solved in Ref. \cite{Andreev 1403 exchange}. This topic is reviewed in Ref. \cite{Andreev 1407 Review} as a part of a detailed description of the many-particle quantum hydrodynamic method. Application of the generalized Coulomb exchange interaction force field was applied to the three-dimensional magnetized quantum plasmas in Ref. \cite{Trukhanova 1405 exchange}. Kinetic model of the low frequency plasma oscillations with the Coulomb exchange interaction in the three-dimensional medium can be found in Ref. \cite{Zamanian PRE 13}. A brief review of development and application of the exchange Coulomb interaction for the three-dimensional electron gas can be found in Ref. \cite{Andreev 1403 exchange}. Reviews of some recent results in quantum plasmas, including relativistic effects in quantum plasmas, can be found in Refs. \cite{Shukla RMP 11}, \cite{Uzdensky arxiv review 14} and introduction of Ref. \cite{bal_eq_2014}.

The spin-1/2 quantum plasmas, for the first time, were considered  by Kuz'menkov et al in 2001 in Ref. \cite{MaksimovTMP 2001}, where corresponding quantum hydrodynamic equations were derived with the explicit consideration of many-particle evolution. The spin-spin interaction between electrons was explicitly included in the derivation. The self-consistent field approximation of the spin-1/2 quantum hydrodynamic equations was considered there. The following paper \cite{MaksimovTMP 2001 b} presented generalization of the spin-1/2 quantum hydrodynamic equations including both the Coulomb exchange interaction and the exchange spin-spin interaction. This model was a generalization of the many-particle quantum hydrodynamics with no explicit spin evolution \cite{LSK1999}, where the general forms of the exchange correlations were presented for both the systems of bosons and the systems of fermions. These general methods were applied to the quantum hydrodynamic derivation of the Gross-Pitaevskii equation, its generalization and similar equations for ultracold fermions for neutral particles with the short-range interaction \cite{Andreev PRA08}. Following Ref. \cite{MaksimovTMP 2001 b} the contribution of the Coulomb and spin-spin exchange interactions in spectrums of magnetized quantum plasmas was studied in Ref. \cite{Andreev AtPhys 08}. In Ref. \cite{Andreev kinetics 13} methods of many-particle quantum hydrodynamics \cite{MaksimovTMP 2001 b}, \cite{MaksimovTMP 2001} were applied to derive the quantum kinetic equations for spinning particles in the three-dimensional and low-dimensional systems. Models for the separate description of the spin-up and spin-down electrons have been developed as well \cite{Harabadze RPJ 04}, \cite{Andreev PRE 15}, \cite{Andreev spin-up and spin-down 1406 Oblique}, \cite{Andreev spin-up and spin-down 1408 2D}.

Due to steady interest to waves on nanotubes we derive the force field of the exchange Coulomb interaction between electrons on small radius cylinders: nanotubes, where curvature of the surface is essential. To this end we consider electron gas with the concentrations $10^{8}\div10^{9}$ cm$^{-2}$ on cylindrical surfaces with radius of $10\div30$ nm.

One can consider spherical and cylindrical waves in the three-dimensional classical \cite{Hossain PP 11} or quantum \cite{Sahu PP 07}-\cite{Hasan ASS 13} plasmas. However there are a lot of papers dedicated to plasmas located on the two-dimensional spherical and cylindrical surfaces. These surfaces exist in the three-dimensional physical space and they can be surrounded by other mediums. As a model problem one can consider spherical or cylindrical surface surrounded by a medium with unitary dielectric permittivity. Hence the surrounding medium does not affect the properties of the plasmas on these surfaces. One can study different properties of plasmas located on these surfaces \cite{Abdikian PP 13}-\cite{Spherical Quant eq state 02}. Speaking of the quantum effects we mean the contribution of the quantum Bohm potential. In spite of the quantum nature of the Fermi pressure existing due to the Pauli exclusion principle it can be included in the classic model by an appropriate choice of the equation of state. Our paper is dedicated to another quantum effect in plasmas. This is the contribution of the exchange part of the Coulomb interaction. For the two-dimensional electron gas it was derived in Ref. \cite{Datta JAP 83} and generalized in Ref. \cite{Andreev 1403 exchange}. Here we consider features of the Pauli pressure and the Coulomb exchange interaction for nanotubes including the curvature of the cylindric surface. Effects of the Coulomb exchange interaction on properties of waves in nanotubes were considered in Refs. \cite{Akbari PP 14}, \cite{Khan JAP 14} for the highly dense electron gas, when the curvature of the cylindric surface is not essential. However, in Ref. \cite{Khan JAP 14}, studying a two dimensional system authors applied the exchange interaction potential derived for three dimensional systems. Let us mention that the Fermi pressure for the electron gas on the spherical surface was recently derived in Ref. \cite{Andreev 1407 Review} (see formula (256)). We should also mention that application of the three-dimensional equations of state or the potentials of the exchange interaction, as it was presented in Refs. \cite{Fathalian SSC 10}, \cite{Fathaliana CPB 12}, \cite{Spherical Quant eq state 01}, \cite{Spherical Quant eq state 02}, \cite{Khan JAP 14} does not appropriate. The models of two-dimensional objects deal with the two-dimensional concentrations: numbers of particle per unit of surface $[n_{2D}]=cm^{-2}$, while the three-dimensional equations of state or the exchange interaction potentials contain the three dimensional concentration $[n_{3D}]=cm^{-3}$. Unfortunately, there are opposite mistakes as well, when one applies an appropriate equation of state for a low-dimensional system, but one considers three-dimensional differential form of the Maxwell equations, which do not reflect geometry of the system under consideration \cite{Manfredi PRB 2001}, \cite{Eq of state Strange diff D}.

Models of quantum plasmas considering contribution of the quantum Bohm potential at huge densities do not include finite size of ions and dust. A hydrodynamic model explicitly including the effects of particle finite size was developed in Ref. \cite{Andreev 1401 finite ions}.

One of most important examples of nanotubes is the carbon nanotube, which is rather different from the objects under consideration. The carbon nanotubes have different regime of the carriers concentration and conductivity. $\sigma$ and $\pi$ electrons in carbon nanotubes reveals different behavior since three $\sigma$ electrons of each atom are strongly bound within the layer of carbon atoms, while one $\pi$ electron of each atom is bound rather weakly. These properties are captured by the two-fluid hydrodynamic model of electrons for carbon nanotubes \cite{Cazaux SSC 70}, \cite{Barton JCP 91}. Conductivity of the bounded electrons involves electrons from each atom, that reveals in the large concentration of carriers $n_{0}\approx10^{15}$ cm$^{-2}$. The fermi pressure for $\sigma$ and $\pi$ electron fluids was introduced in Ref. \cite{Mowbray PRB 04}. Separate spin evolution of spin-up and spin-down electrons in three-dimensional and two-dimensional electron gas has recently been developed in literature \cite{Andreev PRE 15}, \cite{Andreev spin-up and spin-down 1406 Oblique}, \cite{Andreev spin-up and spin-down 1408 2D}.

This paper is organized as follows. In Sec. II the hydrodynamic equations for electron gas on the nanotube are presented and the equation of state for pressure of these electrons is derived. In Sec. III we derive the explicit form of the exchange force field. In Sec. IV we present the spectrum of collective excitations. In Sec. V a brief summary of obtained results is presented.

\section{Method}

\subsection{Basic principles of the many-particle quantum hydrodynamics}

We consider quantum mechanical system of N particles with various masses and charges, interacting by the Coulomb interaction, placed into external electromagnetic field. Derivation of quantum hydrodynamical equations is carried out by the method described in \cite{LSK1999}.
Microscopic number density is defined by formula
\begin{equation}
\label{number_density}
n(\mathbf{r},t)=\int{dR}\sum_{i=1}^N \delta(\mathbf{r}-\mathbf{r}_i)\psi^*(R,t)\psi(R,t),
\end{equation}
where $R=(\mathbf{r}_1,...,\mathbf{r}_N)$, $\mathbf{r}_i$ are the coordinates of $i$th particle,
$dR=\prod_{j=1}^N d\mathbf{r}_j$ is the element of volume in 3N-dimensional configurational space, $d\mathbf{r}_j$ is the element of volume in 3D space of radius-vector $\mathbf{r}_j$. Definition (\ref{number_density}) corresponds the fundamental quantum mechanical definition of a quantum observable value \cite{Landau Vol 3}.

The Hamiltonian of systems under consideration has form
\begin{equation}\label{Hamiltonian}\hat H=\sum_{i=1}^N\left(\frac{\mathbf{D}_i ^2}{2m_i}+e_i \varphi_i\right)+\frac{1}{2}\sum_{i,j=1,i\neq j}^{N}\frac{e_i e_j}{|\mathbf{r}_i-\mathbf{r}_j|},\end{equation}
where $\textbf{D}_{i}=-\imath\hbar\nabla_{i}-e_i\textbf{A}_{i}/c$ is the long momentum including action of the external magnetic field on the charge of the electrons, $\nabla_{i}$ is the derivative over coordinates of $i$-th particle, functions $\varphi_i=\varphi_i(\mathbf{r}_i,t)$, and $\textbf{A}_{i} =\textbf{A}_{i}(\mathbf{r}_{i},t)$ are the potentials of external electromagnetic field. The first group of terms in the Hamiltonian describes charged particles in the external electromagnetic field. The last term in formula (\ref{Hamiltonian}) presents the interparticle Coulomb interaction. The Coulomb interaction gives two contributions in the Euler equation governing the collective motion of electrons: the self-consistent field part and the exchange part. Hamiltonian (\ref{Hamiltonian}) shows that we consider systems of charged particles in the three dimensional space. Below we specify that particles are bound to the cylindrical surface of nanotubes. Our main goal is to present the microscopic derivation of the Coulomb exchange interaction force field of the electrons on the nanotubes.

Differentiating number density (\ref{number_density}) with respect to time and applying the Schr\"{o}dinger equation with the Hamiltonian (\ref{Hamiltonian}), we obtain the continuity equation
\begin{equation}\partial_t n+\nabla (n\textbf{v})=0,\end{equation}
where $\textbf{j}=n\textbf{v}$ is the particle current. Its explicit definition via the many-particle wave function can be found in Refs. \cite{bal_eq_2014}, \cite{Andreev PRA08}, \cite{Andreev PRB 11}, \cite{Andreev RPJ 07}. This definition allows us to derive the Euler equation. The exchange interaction appears beyond the self-consistent field approximation. Hence we need to calculate the quantum two-particle correlations as it was done in Ref. \cite{Andreev 1403 exchange} for three-dimensional and the plane-like two-dimensional electron gas.

\subsection{Set of equations}

The derivation method of the quantum hydrodynamic (QHD) equations for systems of many charged particles was suggested by Kuz'menkov and Maksimov in 1999 in Ref. \cite{LSK1999}. The explicit derivation of the QHD equations in the cylindric coordinates ($\rho=\sqrt{x^{2}+y^{2}}$, $\varphi=\arctan(y/x)$, and $z=z$) has  been recently performed in Ref. \cite{Andreev 1407 Review}. Nevertheless this form of the QHD equations had been applied earlier (see Refs. \cite{Hossain PP 11}-\cite{Hasan ASS 13}). Substituting $\rho=R$, assume $v_{\rho}=0$, and dropping the derivatives over $\rho$ we find QHD equations for electrons on the nanotube
\begin{equation}\label{cont_eq}
\partial_t n+\frac{1}{R}\partial_\varphi(nv_\varphi)+\partial_z(nv_z)=0,
\end{equation}
$$mn\partial_t v_\varphi+mn\biggl(\frac{1}{R}v_{\varphi}\partial_{\varphi}+v_{z}\partial_{z}\biggr)v_{\varphi}+\frac{1}{R}\partial_\varphi P$$
\begin{eqnarray}\label{Euler_fi}
-\frac{\hbar^2}{2m}n\frac{1}{R}\partial_\varphi\biggl(\frac{\triangle\sqrt{n}}{\sqrt{n}}\biggr)=
q_{e}nE_\varphi+F^{Ex}_\varphi,
\end{eqnarray}
and
$$mn\partial_t v_z+mn\biggl(\frac{1}{R}v_{\varphi}\partial_{\varphi}+v_{z}\partial_{z}\biggr)v_{z}+\partial_z P$$
\begin{eqnarray}\label{Euler_z}
-\frac{\hbar^2}{2m}n\partial_z\biggl(\frac{\triangle\sqrt{n}}{\sqrt{n}}\biggr)=
q_{e}nE_z+F^{Ex}_z,
\end{eqnarray}
where $F^{Ex}_{\varphi}$ and $F^{Ex}_{z}$ are the force fields associated with the Coulomb exchange interaction, $P$ is the pressure related to the distribution of particles on different quantum states, $q_{e}=-e$ is the charge of electron. Equation (\ref{cont_eq}) is the continuity equation, and equations (\ref{Euler_fi}), (\ref{Euler_z}) are the Euler equations for two projections of the velocity field $v_\varphi$ and $v_{z}$.

Initially the interaction force field, presented in the Euler equations (\ref{Euler_fi}) and (\ref{Euler_z}) as $F_{\alpha}^{int}=q_{e}nE_{\alpha}+F_{\alpha}^{Ex}$, appears as follows
$$\textbf{F}^{int}=-q_{e}\int d\textbf{r}' (\nabla G(\mid \textbf{r}-\textbf{r}'\mid)) \times$$
\begin{equation}\label{Force General} \times[q_{e}n_{2(ee)}(\textbf{r},\textbf{r}',t)+q_{i}n_{2(ei)}(\textbf{r},\textbf{r}',t)],
\end{equation}
where $n_{2(ea)}(\textbf{r},\textbf{r}',t)$ is the two-particle concentration, with $a=e,i$. Function $n_{2(ei)}(\textbf{r},\textbf{r}',t)$ does not contain the exchange part (or in other words the Fock part of the Hartree-Fock approximation), since electrons and ions are not identical particles. Consequently we have $n_{2(ei)}(\textbf{r},\textbf{r}',t)=n_{e}(\textbf{r},t)n_{i}(\textbf{r}',t)$. Whereas the electron-electron two-particle concentration contains the exchange part. In the limit of weakly interacting electrons we can write \begin{equation}\label{n2}n_{2(ee)}(\textbf{r},\textbf{r}',t)=n_{e}(\textbf{r},t)n_{e}(\textbf{r}',t)-\rho(\textbf{r},\textbf{r}',t)\end{equation}
for system of electrons being in the same spin state. Result (\ref{n2}) appears due to symmetry of the spin part of the many-particle wave function and antisymmetry of the coordinate part of the wave function. What corresponds to the antisymmetry of the full wave function relatively to permutation of identical particles.

The ratio of spin polarization affects the second term in formula (\ref{n2}). If we want to consider the system of partially polarized fermions we need to include that exchange part of the Coulomb interaction gives equal shifts of the energy, but in opposite directions, relatively to the relative direction of spins of these particles. As the first step we consider system of fully spin polarized electrons to present the generalization for partially spin polarized electrons below.

In the case of fermions $\mathbf{F}^{Ex}$ has next form:
\begin{equation}\label{F_exch}
\mathbf{F}^{Ex}=e^2\int d\mathbf{r}^{\prime}\nabla G(\mathbf{r}-\mathbf{r}')|\rho(\mathbf{r},\mathbf{r}',t)|^2.
\end{equation}

We direct the external magnetic field parallel to the cylinder axes $\textbf{B}=B_{0}\textbf{e}_{z}$. In this regime the Lorentz force equals to zero. However, the magnetic field gives contribution via the spin polarisation of electrons, which affects the equation of state and the force of the Coulomb exchange interaction, as we show it below.

There are some concerns \cite{Krishnaswami PRL and arXiv 14}, \cite{Vranjes EPL 12}, \cite{Vranjes EPL 12_repl} about derivation of the hydrodynamic equations from the single particles models and disregards of the area of applicability of hydrodynamic equations. Formulae presented above demonstrate that we have derived our equations from a many-particle model. Therefore appearance of the Fermi pressure and the Coulomb exchange interaction is justified. Area of applicability of our equation is discovered in the text through the derivation of the explicit forms for the Fermi pressure and the exchange interaction. Corresponding limits are discussed below in the paper.

The QHD equations in the cylindrical coordinates were explicitly derived in Ref. \cite{Andreev 1407 Review}, where was discussed the contribution of the quantum part of the inertial forces, which, together with the quantum part of the momentum flux, forms the familiar form of the quantum Bohm potential presented in equations (\ref{Euler_fi}) and (\ref{Euler_z}).

Function $\rho(\mathbf{r},\mathbf{r}',t)$ has the following definition:
\begin{equation}
\rho(\mathbf{r},\mathbf{r}',t)=\sum_{f}n_f \varphi_f^{*}(\mathbf{r},t)\varphi_f(\mathbf{r}',t),
\end{equation}
where $n_f$ is number of particles in the quantum state described by the set of quantum numbers $f$, $\varphi_f(\mathbf{r},t)$ are the wave functions of these states \cite{LSK1999}, \cite{Andreev PRA08}. For our approximate calculation of the exchange Coulomb interaction we apply the wave functions of free particles on the cylindric surface
\begin{equation}\label{model_functions}
\varphi_{p,l}(\mathbf{r},t)=\frac{1}{\sqrt{2\pi RL}}\exp\biggl(-\frac{i}{\hbar}Et\biggr)
\exp(il\varphi+ipz/\hbar),
\end{equation}
where $L$ and $R$ are the length and radius of cylinder (nanotube) correspondingly, $l$ is the quantum number characterizing the momentum over the coordinate $\varphi$ (the spectrum of this number belongs to integers), $p$ is the momentum related to the motion along $z$ axis. The electric field $\mathbf{E}$ in equations (\ref{Euler_fi}), (\ref{Euler_z}) appears, in accordance with the Maxwell equations, as follows
\begin{equation}\label{SUSD2D El field} \textbf{E}=-q_{e}\nabla\int \frac{n_{e}-n_{0i}}{\mid \textbf{r}-\textbf{r}'\mid}d\textbf{r}',\end{equation}
where $d\textbf{r}'$ is the differential of two dimensional surface: $d\textbf{r}'=dx'dy'$ for planes, $\nabla=\{(1/R)\partial_{\varphi}, \partial_{z}\}$ and  $d\textbf{r}'=R d\varphi' dz'$ for cylinders.
We assume that equilibrium concentrations of electrons and ions equal to each other. Hence the difference $n_{e}-n_{0i}$ gives us the perturbation of electron concentration $\delta n$. Equations (\ref{cont_eq})-(\ref{Euler_z}) together with formula (\ref{SUSD2D El field}) correspond to the two-dimensional hydrodynamics developed by Fetter in Refs. \cite{Fetter AoP 73}, \cite{Fetter AoP 74}.

\begin{figure}
\includegraphics[width=8cm,angle=0]{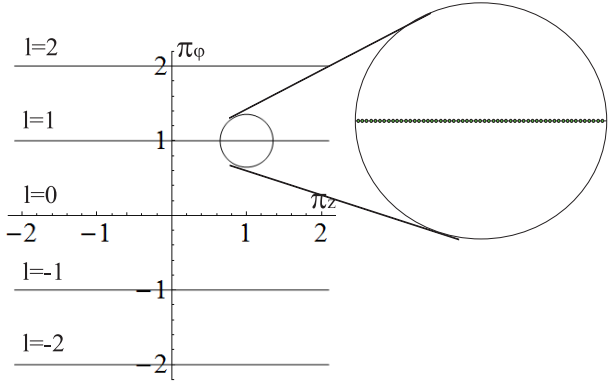}
\caption{\label{BF01} (Color online) The figure shows the distribution of the electronic quantum states on nanotubes in the dimensionless momentum space $\pi_{\varphi}=p_{\varphi}R/\hbar$, $\pi_{z}=p_{z}R/\hbar$. The red dots inside of the large circle show the discrete (quasi-continuous) distribution of the quantum states for a chosen $l$ at different $\pi_{z}$.}
\end{figure}

\begin{figure}
\includegraphics[width=8cm,angle=0]{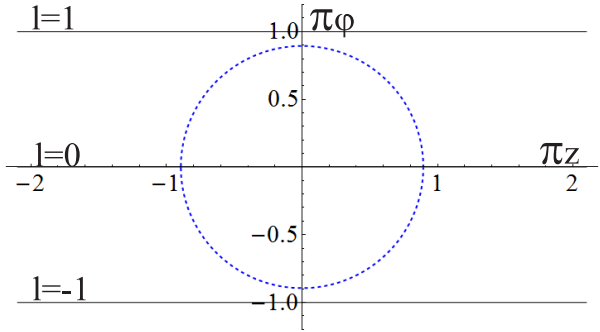}
\caption{\label{BF03} (Color online) The figure shows the two dimensional area in the dimensionless momentum space with the momentum smaller than the momentum $p_{\uparrow\uparrow}$ for electrons on the nanotubes relatively to quantum states with $l=\pm1$. Since all electrons have $l=0$, the distribution of the electron states is the quasi-one-dimensional inspite the two-dimensional distribution of particles in the physical space.}
\end{figure}

Since we consider the electrons on the cylindric surface we can rewrite the general formula (\ref{SUSD2D El field}) in a more explicit form including the symmetry of our problem
$$\textbf{E}=-q_{e}\nabla\sum_{l=-\infty}^{+\infty}\int_{-\infty}^{+\infty}\frac{dk}{(2\pi)^{2}}G(k,l)\times$$
\begin{equation}\label{SUSD2D El field on Cylinder} \times\int R d\varphi' dz' [n_{e}-n_{0i}]\exp[\imath l(\varphi-\varphi')+\imath k(z-z')],\end{equation}
where $G(k,l)=4\pi I_{n}(kR)K_{n}(kR)$, where we have applied the following factorization for the Green function of the Coulomb interaction \cite{Jackson} (see page 104 formula 3.148):
$$\frac{1}{|\mathbf{r}-\mathbf{r}'|}=\frac{1}{\pi}\int^{+\infty}_{-\infty} dk\sum_{l=-\infty}^{+\infty}e^{il(\varphi-\varphi')}\times$$
\begin{equation}
\times e^{ik(z-z')}I_l(k\rho_{<})K_l(k\rho_{>}),
\end{equation}
where $\rho_{<}$, $\rho_{>}$ are the smaller and the larger variable from $\rho$ and $\rho'$, $I_l(k\rho_{<})$ and $K_l(k\rho_{>})$ are modified Bessel functions of the first and second kind. For our case of the electrons located on the cylindrical surface  we use the following limit $\rho=\rho'=R$.

Fig. (\ref{BF01}) shows the distribution of the quantum states in the momentum space for the ideal electron gas located on the cylindrical surface. Fig. (\ref{BF03}) shows the position of the Fermi surface $\mid p\mid\leq p_{Fe}$ relatively to the quantum states of electrons at rather low concentration of electrons.

\begin{figure}
\includegraphics[width=8cm,angle=0]{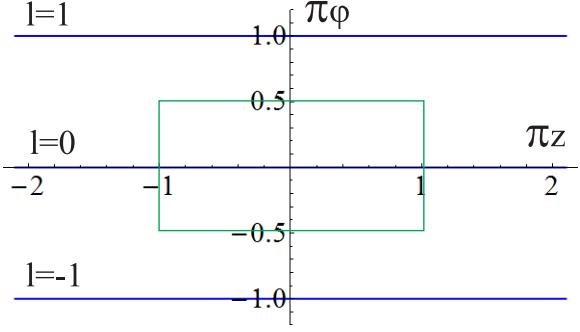}
\caption{\label{BF04} (Color online) The figure shows the volume of the occupied quantum states in the dimensionless momentum space corresponding to the Fermi surface presented in Fig. \ref{BF03}.}
\end{figure}

It can be shown that for sufficiently narrow tube all particles will be in states with $l=0$. This case is considered further. Let us to obtain expressions for Fermi momentum and pressure. Number of particles N in system equals to volume of our system in the momentum space $V=2p_{\uparrow\uparrow} \hbar/R$ divided on the volume of the single electron quantum state on nanotubes $V_{s}=\frac{(2\pi\hbar)^2}{2\pi LR}$ where the volume of the quantum state in the momentum space relatively to the motion in the $\varphi$ direction $V_{\varphi}=2\pi\hbar/(2\pi R)=\hbar/R$. In the limit of full spin polarization, when each quantum state occupied by one electron, we obtain
\begin{equation}
N=\frac{2p_{\uparrow\uparrow} \frac{\hbar}{R}}{\frac{2\pi\hbar^2}{LR}},
\end{equation}
Factor $\hbar/R$ in the numerator arrives from the fact that momentum over coordinate $\varphi$ equals to $p_{\varphi}=l\hbar/R$, from which we see that width of one state is $\hbar/R$ as it depicted in Fig. \ref{BF04}. We obtain that the maximal momentum of occupied states $p_{\uparrow\uparrow}$ of fully polarized electrons on the cylindrical surface equals to
\begin{equation}
p_{\uparrow\uparrow}=2\pi^2\hbar Rn,
\end{equation}
where $n=N/V_{2D}$, $V_{2D}=2\pi LR$.

The Fermi momentum corresponds to double occupation of the low laying quantum states $p_{Fe}=\pi^2\hbar Rn$, which is two times smaller than $p_{\uparrow\uparrow}$. Corresponding Fermi energy has the traditional definition $E_{Fe}=p_{Fe}^{2}/(2m)$.

The average energy of fully spin polarized electrons $\mathcal{E}$ is obtained by formula
$$\mathcal{E}_{\uparrow\uparrow}=\frac{\int_{-p_{\uparrow\uparrow}}^{p_{\uparrow\uparrow}}\frac{1}{2m}p_{z}^2dp_{z}
\int_{-\hbar/2R}^{\hbar/2R}dp_{\varphi}}{\int_{-p_{\uparrow\uparrow}}^{p_{\uparrow\uparrow}}dp_{z}
\int_{-\hbar/2R}^{\hbar/2R}dp_{\varphi}}$$
\begin{equation}
=\frac{2\pi^4}{3m}\hbar^2R^2n^2=\frac{4}{3}E_{Fe}.
\end{equation}

Curvature of the cylindrical surface with radius of order of $20$ nm gives contribution in the Fermi pressure and the Coulomb exchange interaction if the concentration of electrons is of order of $10^{8}\div10^{10}$ cm$^{-2}$. It corresponds to the Fermi temperature of order of $T_{Fe}=\frac{\mathcal{E}_{Fe}}{k_{B}}=1$ K, where we apply the reduced Planck constant $\hbar=1.05$ $10^{-27}$ erg$\cdot$s, the Boltzmann constant $k_{B}=1.38$ $10^{-16}$ erg$\cdot$K$^{-1}$, and the electron mass $m_{e}=9$ $10^{-28}$ g. Therefore our results corresponds to low temperatures.

The average energy allows us to find the equation of state. Pressure of the ideal gas of particles having two degrees of freedom arises as
\begin{equation}\label{pV}
P_{\uparrow\uparrow}=\mathcal{E}_{\uparrow\uparrow}n.
\end{equation}
In 3D case in formula (\ref{pV}) coefficient $2/3$ is placed before $\mathcal{E}$, in 2D case coefficient equals to $2/2=1$. As a result next expression is obtained for the pressure of the spin polarized electrons:
\begin{equation}\label{Fermi_pressure}
P_{\uparrow\uparrow}=\frac{2\pi^4}{3m}\hbar^2 R^2n^3.
\end{equation}

%%%%%%%%%%%%%%%%%%%%%%%%%%%%

Similar calculations for unpolarized electrons gives us the Fermi pressure $P_{Fe}$:
\begin{equation}\label{Fermi_pressure unpolarized}
P_{Fe}=\frac{\pi^4}{6m}\hbar^2 R^2n^3.
\end{equation}

%%%%%%%%%%%%%%%%%%%%%%%%%%%%%
\begin{figure}
\includegraphics[width=8cm,angle=0]{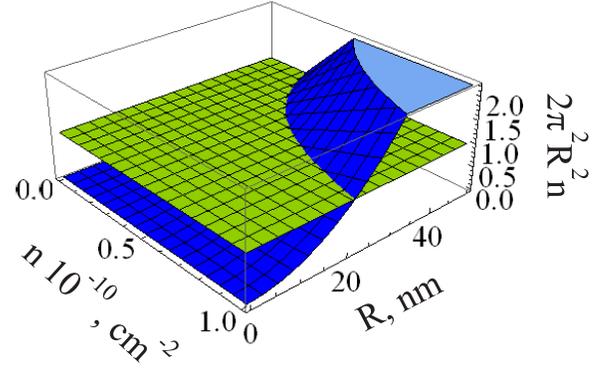}
\caption{\label{BF07} (Color online) The figure shows the conditions for occupation of one level (l=0) only. The single level is occupied if the blue surface is below the plane presenting the unitary level.}
\end{figure}

%%%%%%%%%%%%%%%%%%%%%%%%%%%%

If one considers the plane-like 2DEG he finds the following equation of state $P_{Fe,Pl}=\pi\hbar^{2}n^{2}/(2m)$.

We consider occupation of the single level in the momentum space with $l=0$. Area of parameters corresponding to this regime is presented in Fig. (\ref{BF07})

\section{Force field of exchange interaction}

In this section we present the calculation of force field of the Coulomb exchange interaction (\ref{F_exch}) for electrons located on the cylinder surface. Calculations is presented for fully spin polarized gas: $n_f=1$.

We can rewrite formula (\ref{F_exch}) as follows
$$\mathbf{F}^{Ex}=e^2\int d\mathbf{r}^{\prime}\nabla G(\mathbf{r}-\mathbf{r}')\times$$
\begin{equation}\label{F_exch2}
\times\sum_{f,f'}n_f n_{f'}\varphi_f^{*}(\mathbf{r},t)\varphi_f(\mathbf{r}',t)
\varphi_{f'}(\mathbf{r},t)\varphi_{f'}^{*}(\mathbf{r}',t).
\end{equation}
As mentioned above, the occupation numbers for states inside the sphere of radius $p_{\uparrow\uparrow}$ in the momentum states equal to one: $n_f=1$. Let us substitute wave functions (\ref{model_functions}) in formula (\ref{F_exch2}). Sum over states appears as
\begin{equation}
\sum_f=\frac{1}{2\pi\frac{\hbar}{L}}\int_{-p_{\uparrow\uparrow}}^{p_{\uparrow\uparrow}}dp.
\end{equation}
We do not need to perform the summation over all $l$ since in our case $l=0$ for all particles. As a result next expression for $\mathbf{F}^{Ex}$ is obtained:
$$\mathbf{F}^{Ex}=\frac{e^2}{4\pi^3 \hbar^{2} R}\times$$
\begin{equation}\label{Exchange force l=0}
\times\nabla\int_{-p_{\uparrow\uparrow}}^{p_{\uparrow\uparrow}}dp\int_{-p_{\uparrow\uparrow}}^{p_{\uparrow\uparrow}}dp'
I_0((p-p')R/\hbar)K_0((p-p')R/\hbar).
\end{equation}
%{\color{red}smf like $L^2$ should be in the denominator or $n$ in the numerator since we have force DENSITY!!}

We consider the limit of small wave vectors $kR\ll1$, therefore, the asymptotic expressions of the modified Bessel functions can be used for calculation of the exchange interaction. We apply
\begin{equation}
I_0(x)\simeq 1,
\end{equation}
and
\begin{equation}
K_0(x)\simeq -\ln\frac{|x|}{2}-\gamma,
\end{equation}
where $\gamma\approx0,577215...$ is the Euler-Mascheroni constant (see \cite{Jackson} page 92 formulae 3.102 and 3.103). As a result for the force field density next expression is obtained:
\begin{equation}\label{Exchange force l=0 small k explicit}
\mathbf{F}^{Ex}=2\pi e^2 R\nabla\biggl[\biggl(3-2\gamma-2\ln [2\pi^2R^2n]\biggr)n^2\biggr],
\end{equation}
with $\ln [2\pi^2R^2n]<0$.

\begin{figure}
\includegraphics[width=8cm,angle=0]{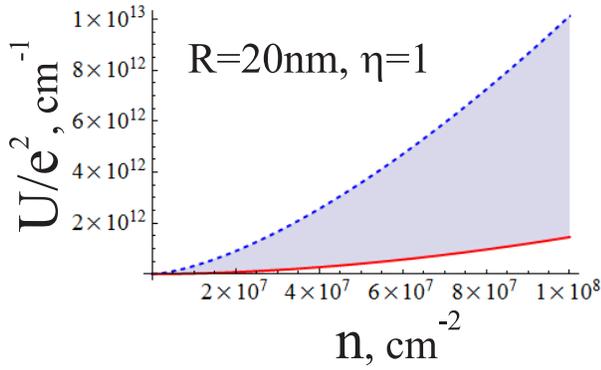}
\caption{\label{BF08} (Color online) The figure shows the comparison of the potential of the exchange forces for the plane-like electron gas and the electron gas on the nanotube with the occupation of one level $l=0$. The potential corresponds to the force field as $\textbf{F}^{Ex}=\nabla U$. The upper (blue, dashed) curve describes the exchange interaction in the plane-like electron gas. The lower (red) curve presents the exchange interaction in the cylindrical two-dimensional electron gas.}
\end{figure}

It is interesting and important to compare result (\ref{Exchange force l=0 small k explicit}) with the two-dimensional
analog $\textbf{F}_{pl}^{Ex}=192\cdot \textrm{arcsh}1\cdot e^{2}\nabla n^{3/2}/(3\pi\sqrt{\pi})$ \cite{Datta JAP 83},\cite{Andreev 1403 exchange}. To present the comparison it is rather useful
to introduce the potential of these forces $\mathbf{F}^{Ex}=\nabla U_{Cyl}$, and $\textbf{F}_{pl}^{Ex}=\nabla U_{pl}$. Dependence of the potentials on the concentration $U_{Cyl}(n)$, $U_{pl}(n)$ are presented in Fig. \ref{BF08}.

%ExCyl__Exchange

\begin{figure}
\includegraphics[width=8cm,angle=0]{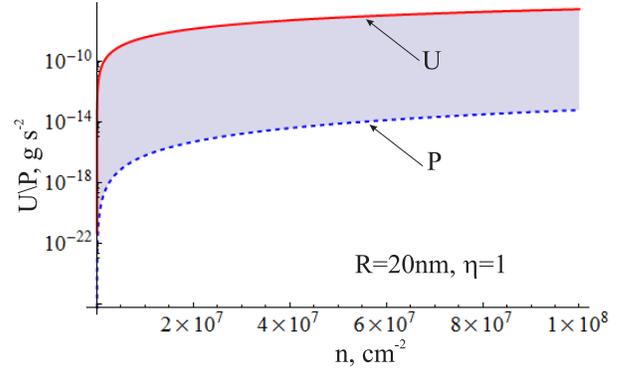}
\caption{\label{BF13} (Color online) The figure shows the Fermi pressure $P$ given by formula (\ref{Fermi_pressure intermediate Spin}) (the blue dashed line) and the Coulomb exchange interaction potential $U$ given by formula (\ref{Exchange force Partial Polarisation}) (continues red line) for the fully spin polarized systems.}
\end{figure}

\begin{figure}
\includegraphics[width=8cm,angle=0]{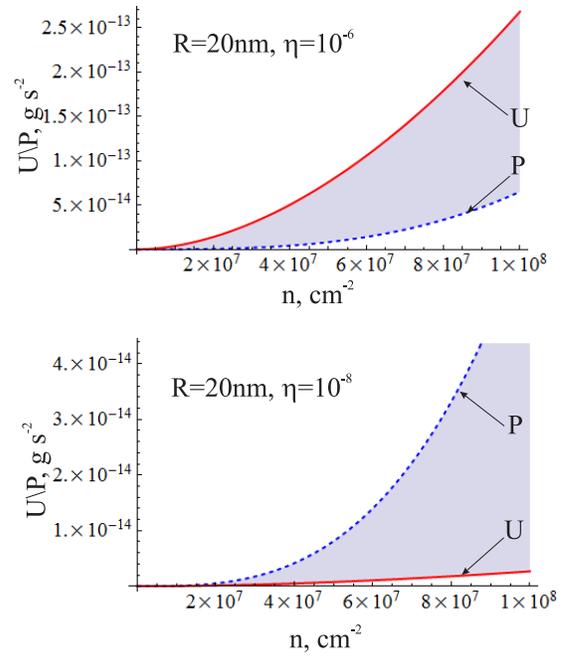}
\caption{\label{BF14} (Color online) The figure shows the Fermi pressure $P$ given by formula (\ref{Fermi_pressure intermediate Spin}) (the blue dashed line) and the Coulomb exchange interaction potential $U$ given by formula (\ref{Exchange force Partial Polarisation}) (continues red line) for two regimes of the partially spin polarized electrons.}
\end{figure}

\subsection{Partial spin polarisation of electrons}

Above we have presented the equation of state (\ref{Fermi_pressure}) and the force field of the Coulomb exchange interaction (\ref{Exchange force l=0 small k explicit}) for electrons on nanotubes in the case of the full spin polarisation. This is a simple case, which allows to describe the derivation in a simple way. Now we are going to realistic regime of the \emph{partial spin polarisation}.

The potential of the exchange interaction is significantly larger than the Fermi pressure in wide range of the spin polarizations $\eta\in[10^{-5}, 1]$. We illustrate it by Fig. (\ref{BF13}), where we compare the exchange interaction potential and the Fermi pressure for the electron gas in the regime of the full spin polarization $\eta=1$. We see that the exchange interaction potential is five orders larger than the Fermi pressure.

Fig. (\ref{BF14}), similarly to Fig. (\ref{BF13}), gives a comparison of the exchange interaction potential $U$ (\ref{Exchange force Partial Polarisation}) and the Fermi pressure $P$ (\ref{Fermi_pressure intermediate Spin}) in the regime of low spin polarization $\eta=10^{-6}$ and $\eta=10^{-8}$. In this regime $P$ and $U$ are comparable. The Fermi pressure becomes larger than the the exchange interaction potential $U$ at small enough spin polarization.

At partial spin polarization of degenerate electrons we have that the spin-down (the spin-up) electrons occupy the quantum states laying inside the circle with radius $p_{\downarrow\downarrow}$ ($p_{\uparrow\uparrow}$) in the momentum space, and $p_{\downarrow\downarrow}>p_{\uparrow\uparrow}$. Hence the area at $p<p_{\downarrow\downarrow}$ is occupied by pair of electrons with opposite spins, and the area $p_{\downarrow\downarrow}>p>p_{\uparrow\uparrow}$ is occupied by the spin-down electrons only. The full pressure of the system $P$ appears as the sum of the partial pressures of spin-up and spin-down electrons $P=P_{\uparrow\uparrow}+P_{\downarrow\downarrow}$, where the partial pressures $P_{\uparrow\uparrow}$, $P_{\downarrow\downarrow}$ are given by formula (\ref{Fermi_pressure}) with the concentration of the spin-up electrons $n_{\uparrow}$ and the spin-down electrons $n_{\downarrow}$ correspondingly. The partial concentrations $n_{\uparrow}$ and $n_{\downarrow}$ can be easily presented via the full concentration of electrons and the ratio of the spin polarization
\begin{equation}\begin{array}{cc}
                  n_{\uparrow}=\frac{n}{2}-\frac{\Delta n}{2}, & n_{\downarrow}=\frac{n}{2}+\frac{\Delta n}{2},
                \end{array}
\end{equation}
where $n$ is the full concentration, and $\Delta n$ is the concentration of particles being in the partially occupied states at $p_{\downarrow\downarrow}>p>p_{\uparrow\uparrow}$. Applying the notion of spin polarization ratio $\eta=\mid n_{\uparrow}-n_{\downarrow}\mid/(n_{\uparrow}+n_{\downarrow})$. Therefore $n_{\uparrow}=n(1-\eta)/2$, and $n_{\downarrow}=n(1+\eta)/2$.

%%%%%%%%%%%%

\begin{figure}
\includegraphics[width=8cm,angle=0]{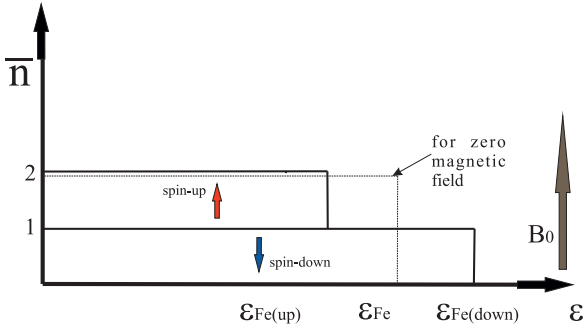}
\caption{\label{BF11} (Color online) The figure shows the distribution functions for the partially spin polarized electrons and non-polarized electrons.}
\end{figure}

Finally, for the case of the intermediate spin polarization we have
\begin{equation}\label{Fermi_pressure intermediate Spin}
P_{\Updownarrow}=(1+3\eta^{2})\frac{\pi^4}{6m}\hbar^2 R^2n^3.
\end{equation}

Fig. (\ref{BF11}) shows the Fermi step (the distribution function) for the partially polarized and the non-polarized systems of electrons.

Next we need to consider the force field of the exchange Coulomb interaction for the partially spin polarized electrons.

Due to partial spin polarization at temperatures below the Fermi temperature we obtain
$$F^{(Ex)}=F^{(Ex)}_{\downarrow\downarrow}-F^{(Ex)}_{\uparrow\uparrow}$$
$$=2\pi e^2 R\nabla\biggl[\biggl(3-2\gamma-2\ln [2\pi^2R^2n_{\downarrow}]\biggr)n_{\downarrow}^2$$
$$-\biggl(3-2\gamma-2\ln [2\pi^2R^2n_{\uparrow}]\biggr)n_{\uparrow}^2\biggr]$$
$$=2\pi e^2 R\nabla\biggl[(3-2\gamma)\eta n^{2}+\frac{1}{2}n^{2}(1+\eta^{2})\ln\biggl(\frac{1-\eta}{1+\eta}\biggr)$$
\begin{equation}\label{Exchange force Partial Polarisation} -n^{2}\eta\biggl(2\ln[\pi^{2}R^{2}n]+\ln(1-\eta^{2})\biggr)\biggr].\end{equation}

If we have a set of quantum states occupied by two electrons with opposite spins, we find that a chosen fermion interact  with almost equal numbers of the spin-up and spin-down electrons. Hence its exchange interaction with the surrounding electrons equal to zero. Therefore we find that the exchange interaction is related to the part of electrons being in states occupied by a single electron. In Fig. (\ref{BF09}) it corresponds to electrons in states located between the circles.

\section{Dispersion equation}

%%%%%%%%%%%%
\begin{figure}
\includegraphics[width=8cm,angle=0]{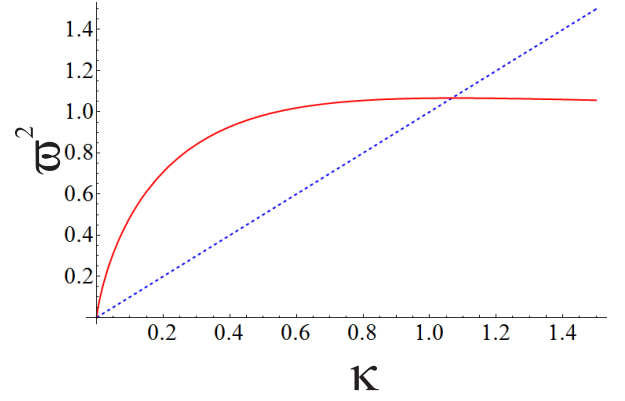}
\caption{\label{BF10} (Color online) The figure shows the $\varpi^{2}=\omega^{2}_{Le,Cyl}/(2\pi e^{2}n_{0}^{\frac{3}{2}}/m)=\frac{1}{2\pi}\kappa G(\sqrt{n_{0}}R\cdot\kappa)$ by red (continues) curve presenting the electron gas of the nanotube, and $\varpi^{2}=\omega^{2}_{Le,Pl}/(2\pi e^{2}n_{0}^{\frac{3}{2}}/m)=\kappa$ for blue (dashed) line presenting the Langmuir frequency of the plane-like two-dimensional electron gas and $\kappa=k/\sqrt{n_{0}}$.}
\end{figure}

Using equations obtained above, let us investigate dispersion for the linear plasma waves in the electron gas on nanotubes. Let us present the concentration of particles and the velocity field as $n=n_0+\delta n$, $v_{\varphi}=0+\delta v_{\varphi}$, and $v_{z}=0+\delta v_{z}$, where $n_0$ is the equilibrium value of concentration, $\delta n$ is the perturbation of the concentration due to wave propagation. We assume that there are no currents in the equilibrium. Now let us apply the Fourier transformation to linearized form of equations (\ref{cont_eq})-(\ref{Euler_z}) in accordance with formula (\ref{model_functions})
\begin{equation}\label{cont eq lin}
-\omega \delta n+k\delta v_z=0,
\end{equation}
\begin{equation}\label{Euler eq varphi lin}
\omega \delta v_\varphi=0,
\end{equation}
and
$$mn_0\omega \delta v_z-(1+3\eta^{2})\frac{2\pi^4}{m}\hbar^2 R^2n_0^2k \delta n$$
$$ -\frac{\hbar^2}{4m}k^3\delta n=e^2n_0 G(k)\delta n
$$
$$+ 4\pi e^2R n_0k
\biggl[2(1-\gamma)\eta+\frac{1}{2}(1+\eta^{2})\ln\biggl(\frac{1-\eta}{1+\eta}\biggr)$$
\begin{equation}\label{Euler eq z lin}
-\eta[\ln(1-\eta^{2})+2\ln(\pi^2n_0R^2)]\biggr]\delta n.
\end{equation}
At derivation of equations (\ref{cont eq lin})-(\ref{Euler eq z lin}) we have included that derivatives
on the angle $\varphi$ equal to zero since we have $l=0$.

Equations (\ref{cont eq lin})-(\ref{Euler eq z lin}) allows to obtain the dispersion dependence of
the Langmuir waves on the cylindrical surface of the nanotubes
$$\omega^2(k)=\frac{e^2n_0k}{m}G(k)+(1+3\eta^{2})\frac{2\pi^4}{m^2}\hbar^2 R^2 n_0^2 k^2
$$
$$-\frac{4\pi e^2}{m} R n_0k^2
\biggl[2(1-\gamma)\eta+\frac{1}{2}(1+\eta^{2})\ln\biggl(\frac{1-\eta}{1+\eta}\biggr)$$
\begin{equation}\label{Dispersion dependence partial spin polarisation}
-\eta[\ln(1-\eta^{2})+2\ln(2\pi^2n_0R^2)]\biggr].
\end{equation}

At zero spin polarization $\eta=0$ from formula (\ref{Dispersion dependence partial spin polarisation}) we find
\begin{equation}\label{Dispersion dependence zero spin polarisation}
\omega^2(k)=\frac{e^2n_0k}{m}G(k)+\frac{2\pi^4}{m^2}\hbar^2 R^2 n_0^2 k^2.
\end{equation}

In the opposite limit, the full spin polarization $\eta=1$ formula (\ref{Dispersion dependence partial spin polarisation}) gives
$$\omega^2(k)=\frac{e^2n_0k}{m}G(k)+\frac{8\pi^4}{m^2}\hbar^2 R^2 n_0^2 k^2
$$
\begin{equation}\label{Dispersion dependence full polarisation}
-\frac{8\pi e^2}{m} R n_0 k^2
[1-\gamma-\ln(2\pi^2n_0R^2)].
\end{equation}
with $1-\gamma-\ln(2\pi^2n_0R^2)>0$, since $\ln(2\pi^2n_0R^2)<0$.

In formula (\ref{Dispersion dependence full polarisation}) we have used $$\frac{1}{2}(1+\eta^{2})\ln\biggl(\frac{1-\eta}{1+\eta}\biggr)-\eta[\ln(1-\eta^{2})-2\ln(\pi^2n_0R^2)]$$
$$=\ln(1-\eta)-\ln(1-\eta)-2\ln2-2\ln(\pi^2n_0R^2)$$
$$=-2\ln2-2\ln(\pi^2n_0R^2)=-2\ln(2\pi^2n_0R^2).$$

The first term in the dispersion dependence (\ref{Dispersion dependence full polarisation}) corresponds to the plasma frequency $\omega_{Le,Cyl}^2=e^2n_0k G(k)/m$. The Fourier image of the Green function of the potential of Coulomb interaction on the cylindrical surface in the three-dimensional space is obtained as $G(k)=4\pi I_0(kR)K_0(kR)$. This function is the signature of geometrical properties of our system. In plane-like two-dimensional electron gas one can find $\omega_{Le,Pl}^{2}=2\pi e^2 n_0k/m$. All of these Langmuir frequencies are analogs of well-known expression for three dimensional medium $\omega_{Le,3D}^{2}=4\pi e^2 n_{0,3D}/m$, where $n_{0,3D}$ is the three dimensional concentration of particles $[n_{0,3D}]=cm^{-3}$. The second term arises from the Fermi pressure (\ref{Fermi_pressure}), and the third term presents the quantum Bohm potential. The last term in the dispersion dependence arises from the exchange interaction. All terms, except the last one, are positive, while the exchange interaction gives a negative contribution decreasing the Langmuir frequency. In spite the quantum nature of the exchange interaction, the force field (\ref{Exchange force l=0 small k explicit}) and its contribution in the dispersion dependence (\ref{Dispersion dependence full polarisation}) does not contain the Planck constant.

Numerical comparison between the Fermi pressure and the exchange interaction (see Fig. (\ref{BF13})) shows that we can drop contribution of the Fermi pressure in compare with the Coulomb exchange interaction in our regime.

Fig. (\ref{BF10}) demonstrates difference of the Langmuir frequencies for the plane-like electron gas and the cylindrical electron gas at $R=20$ nm.

\section{Conclusion}

Nanotubes with radiuses in interval from $10$ nm to $30$ nm containing free charge carriers at concentration $10^{9}$ cm$^{-2}$ and smaller reveal large difference of properties from the charge carriers on the plane-like two dimensional structures. To obtain the quantum hydrodynamic description of these objects we have derived corresponding equation of state for the pressure of degenerate electrons. We have included dependence of the pressure on the spin polarization. Zero spin polarization gives the Fermi pressure for electrons on the nanotubes, which appears to be proportional to the third degree of the electron two-dimensional concentration. We have also performed the explicit derivation of the Coulomb exchange interaction force field. The force field is a potential field, and its potential is proportional to the square of concentration multiplied by the sum of a constant and the logarithm of the concentration. Dependence of the exchange interaction force field on the spin polarization is derived. It has been shown that the force field monotonically increases with the increase of the spin polarization. Numerical analysis of the difference of the pressures and the exchange interaction forces between plane-like electron gas and the electron gas on the nanotubes have been presented.

We have applied obtained model to study the dispersion of collective excitations of charge carriers on the nanotubes assuming that the external magnetic field is directed parallel to the tube axes. We have considered dispersion of the Langmuir waves obtaining contribution of the Fermi pressure, the exchange Coulomb interaction, and the ratio of the spin polarization.

Finally we should note that we have developed a new set of closed hydrodynamic equations for nanotubes with the electron concentration of order $10^{10}$ cm$^{-2}$ and smaller. Our model allows to study different plasma-wave phenomena in these objects.

\begin{acknowledgements}
The authors thank Professor L. S. Kuz'menkov for fruitful discussions. P.A. thanks the Dynasty foundation.
\end{acknowledgements}

\end{document}